\documentclass[twocolumn,superscriptaddress,prl,normalem,nofootinbib]{revtex4-2}
\usepackage{color}
\usepackage{graphicx}
\usepackage{amsmath}
\usepackage{xspace}
\usepackage[normalem]{ulem}
\usepackage{appendix}

\begin{document}
\newcommand{\levele}{$^3H_4$\xspace}
\newcommand{\levelh}{$^3H_5$\xspace}
\newcommand{\levelm}{$^3F_4$\xspace}
\newcommand{\levelg}{$^3H_6$\xspace}
\newcommand{\Fpo}{F_{\mathrm{po}}}
\newcommand{\Fth}{F_{\mathrm{th}}}
\newcommand{\SFFth}{S_{\mathrm{FF}}^{\mathrm{th}}}
\newcommand{\SFFpo}{S_{\mathrm{FF}}^{\mathrm{po}}}
\newcommand{\Deltapo}{\Delta_{\mathrm{po}}}
\newcommand{\Gstress}{G_{\mathrm{stress}}}

\newcommand{\SFFthres}{S_{\mathrm{FF}}^{\mathrm{th,res}}}
\newcommand{\perroothz}{\,\mathrm{Hz}^{-1/2}}
\newcommand{\nact}{n_{\mathrm{active}}}

\newcommand{\TC}[1]{{\color{green}{TC: #1}}}
\newcommand{\ALC}[2]{{\color{cyan}{\textbf{ALC: #1}} #2}}
\newcommand{\PV}[2]{{\color{magenta}{\textbf{PV: #1}} #2}}

\title{Piezo-orbital backaction force in a rare-earth doped crystal} 

\author{A. Louchet-Chauvet}
\email{anne.louchet-chauvet@espci.fr}
\affiliation{ESPCI Paris, Université PSL, CNRS, Institut Langevin, 75005 Paris, France}
\affiliation{Université Paris-Saclay, CNRS, Laboratoire Aim\'e Cotton, 91405 Orsay, France}

\author{P. Verlot}
\affiliation{Université Paris-Saclay, CNRS, ENS Paris-Saclay, CentraleSupélec, LuMIn, 91405, Orsay,
France}
\affiliation{Institut Universitaire de France, 1 rue Descartes, 75231 Paris, France}

\author{J.-P. Poizat}
\affiliation{Univ. Grenoble Alpes, CNRS, Grenoble INP, Institut Néel, 38000 Grenoble, France}

\author{T. Chanelière}
\affiliation{Univ. Grenoble Alpes, CNRS, Grenoble INP, Institut Néel, 38000 Grenoble, France}


\begin{abstract}
We investigate a system composed of an ensemble of room temperature rare-earth ions embedded in a bulk crystal, intrinsically coupled to internal strain via their sensitivity to the surrounding crystal field. We evidence the generation of a mechanical response under resonant atomic excitation. We find this motion to be the sum of two fundamental, resonant optomechanical backaction processes: a conservative, piezo-orbital mechanism, resulting from the modification of the crystal field associated with the promotion of the ions to their excited state, and a dissipative, non-radiative photothermal process related to the phonons generated throughout the atomic population relaxation.
Our work opens new research avenues in hybrid optomechanics, and highlights new interactions that may be key for understanding the dephasing dynamics of ultra-coherent rare-earth ions.
\end{abstract}
\maketitle

\section{Introduction}

Hybrid mechanical systems consist of a mechanical oscillator coupled to a quantum degree of freedom. They have been identified as a promising platform to prepare, detect and manipulate non-classical macroscopic states of mechanical motion~\cite{treutlein2014hybrid}. Amongst various approaches~\cite{arcizet2011NV,viennot2018phonon}, strain-induced coupling is particularly attractive because it intrinsically enables the design of monolithic devices exhibiting mechanical and thermal stability together with scalability~\cite{yeo2014strain}.

Various hybrid designs based on strain coupling have been proposed, where the quantum emitter is either a single system, namely a quantum dot (QD)~\cite{yeo2014strain}, a nitrogen-vacancy (NV) center in diamond~\cite{teissier2014strain}, or an ensemble of rare-earth ions in a crystal (REIC)~\cite{molmer2016dispersive,ohta2021rareearth}.
For QDs, the deformation of the crystal lattice induced by the applied strain results in a change of the semiconductor bandgap which directly defines the energy levels of the QD. Conversely, in impurity-doped systems, the coupling results from the sensitivity of the impurity's electronic orbitals to the crystal field, which is affected by the deformation of the crystalline matrix~\cite{liu2005electronic}. Because these mechanisms are fundamentally different, we will further refer to these strain-coupling mechanisms with suitable terminology, namely \emph{piezo-excitonic} coupling for QDs and \emph{piezo-orbital} coupling for doped crystals.

The first step towards controlling the optomechanical interaction in a strain-coupled hybrid system is to investigate the sensitivity of its quantum degree of freedom to an applied strain by measuring the induced detuning of the atomic resonance~\cite{tumanov2018static,galland2020mechanical,doherty2014electronic}. This mechanism offers interesting applications such as strain-based sensing in bulk materials~\cite{he1995determination,rajendran2017method,louchet2019piezospectroscopic}, microstructures~\cite{deassis2017strain} and heterostructures~\cite{schulein2015fourier}.

Further control of a hybrid optomechanical system requires exploring the associated  backaction effect, \emph{ie} generating strain or motion by addressing the quantum states of the atomic system, using an optical excitation. In REIC, this backaction stems from the alteration of the ion's electronic orbital following a change of state. This affects the crystal field around the excited ion, which results in a local rearrangement of the host matrix, therefore inducing strain.
From an optomechanical point of view, an important distinction between this piezo-orbital backaction and the recently reported piezo-excitonic backaction~\cite{kettler2021inducing} relates to their associated interaction times:  with typical timescales set by a population lifetime in the millisecond range (to be compared to $\sim 1$~ns for solid state QDs), the piezo-orbital backaction in REIC could give access to highly attractive regimes, including resolved sideband \cite{treutlein2014hybrid}, strong/ultra-strong coupling \cite{aspelmeyer2014cavity}, and reversed dissipation \cite{ohta2021rareearth}.

In this work, we demonstrate the generation of mechanical motion following resonant atomic driving in a strain-coupled hybrid mechanical system based on a large ensemble of ions embedded in a cm-scale monocrystal. The absence of sharp mechanical resonances in such a bulk system enables full temporal and spatial reconstruction of the optically induced motion.
We present an ultra-sensitive, time-resolved,  tomography setup allowing us to separately address the dissipative and conservative backaction contributions.
Remarkably, the dissipative component is fully described by the non-radiative relaxation dynamics of the ions, whereas the conservative contribution is caused by the piezo-orbital effect.
Overall the degree of understanding and control of both the conservative and dissipative part of the optomechanical backaction confirms the potential of rare-earth-based systems as hybrid optomechanical systems.

\section{Optomechanical transduction in Tm:YAG}

In REICs, the optical transitions within the $4f$ configuration are shielded from environment perturbations by the outer $5s$ and $5p$ electrons. Due to this shielding, the rare-earth ions can exhibit very long population and coherence lifetimes~\cite{thiel2011rare}, representing a unique resource for quantum technologies~\cite{usmani2010mapping,fernandez2015coherent,thorpe2011frequency}. Amongst the many REICs available, Tm:YAG shows moderate sensitivity to magnetic and electric fields~\cite{goldner2006hyperfine,minnegaliev2021linear}, making it a good candidate to probe the piezo-orbital dynamics.

\begin{figure}[t]
\centering
\includegraphics[width=8.0cm]{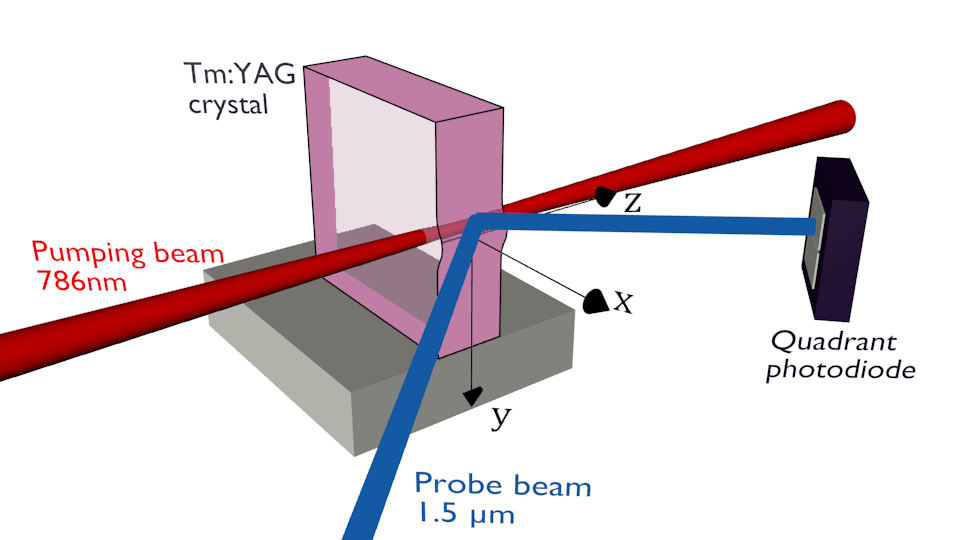}
\caption{Photodeflection setup geometry for evidencing an optomechanical backaction in a bulk crystal. A pump beam is focused in the crystal, propagating close to one of its surfaces. A probe beam is reflected on this surface with a $45^\circ$ angle of incidence. In the presence of an optomechanical transduction, the surface is distorted, leading to a vertical deflection of the reflected probe beam, captured by a quadrant photodetector.}
\label{fig:photodeflection}
\end{figure}

\begin{figure}
  \centering
  \includegraphics[width=8.6cm]{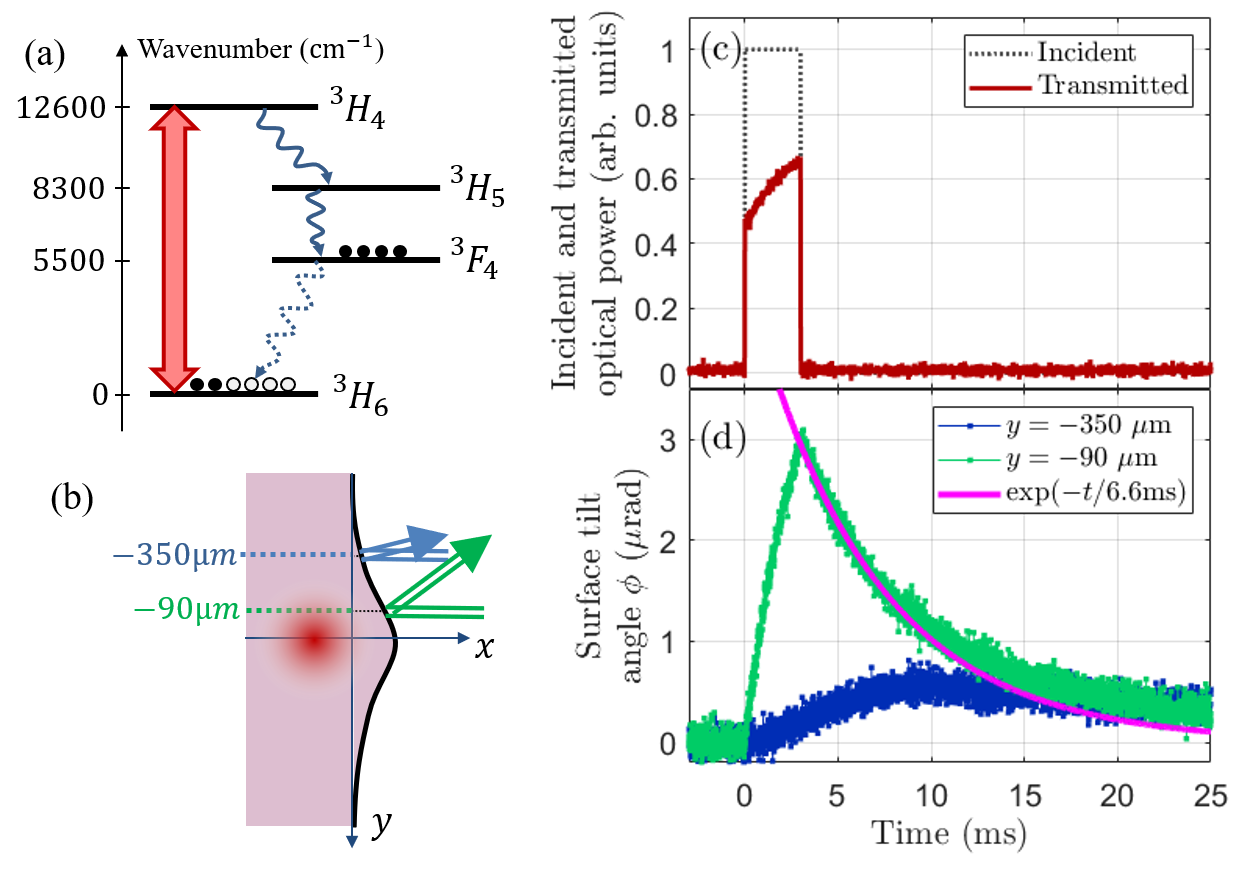}
  \caption{(a) Simplified level system for Tm$^{3+}$ ions. At room temperature, the resonant excitation (red double arrow) along the \levelg$\rightarrow$\levele transition allows a sizeable fraction of the ions in the pump beam volume to be stored in the \levelm level for several ms, via optical pumping. (b) Lateral, schematic view of the pump and probe beam geometry. (c) Incident and transmitted pump laser power at $786$~nm through the Tm:YAG crystal. Optical pumping to \levelm  leads to a progressive increase in the transmitted light.
  (d) Measured surface tilt angle  (averaged 128 times) for two relative positions of the pump and probe beam [schematically shown in (b)]. An exponential decay is  shown as a guide to the eye.}
  \label{fig:unecourbe}
\end{figure}

We consider the geometry depicted in Figure~\ref{fig:photodeflection}. A pump beam shines a cm-scale cuboid Tm-doped YAG crystal close to one of its surfaces, such that the resulting strain field is mainly directed towards this surface and maximizes the corresponding time- and position-dependent surface deformation.
The latter is retrieved from the angular deflection of an auxiliary continuous wave (cw) probe beam reflected on the moving surface. The optical pumping process in the Tm$^{3+}$ ion ensemble is shown in Figure~\ref{fig:unecourbe}(a,c). We estimate that about $N=4\cdot 10^{14}$ ions are optically pumped to the \levelm state after $3$~ms (see SI).

As for a majority of quantum emitters, the relaxation process in Tm:YAG involves both radiative and non-radiative contributions, that will lead to the conservative and dissipative components of the optomechanical backaction process, respectively. The former is the aforementioned piezo-orbital backaction, that is the generation of strain resulting from the population transfer amongst different electronic states, whereas the latter results from non-radiative decay mechanisms, and is also known as the photothermal effect. This effect has been studied in the literature and triggered on purpose to elucidate the non-radiative processes in various REICs~\cite{rodriguez1993simultaneous, grinberg2000photoacoustic}.
Disentangling the two backaction contributions is a challenging problem since they are both optically resonant. In this work we take advantage of the  mechanically nonresonant nature of our bulk sample and reconstruct its time-dependent deformation in response to a rectangular light pulse. Because of the distinct time and space signatures associated with the piezo-orbital and photothermal drives, the time-resolved mapping of the optomechanical backaction force enables an unambiguous determination of each of these contributions.

\section{Experimental results}
\label{sec:experimental}

\begin{figure*}[t!]
\centering
\includegraphics[width=17cm]{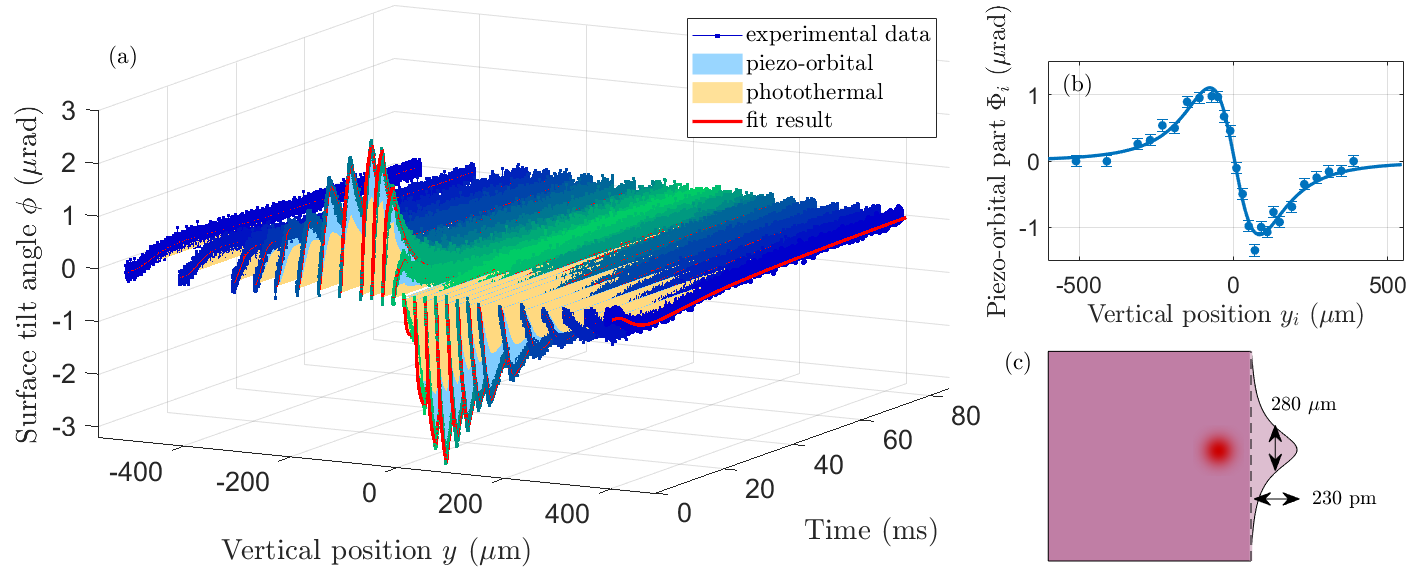}
\caption{Time-resolved tomography of the crystal surface when a $3$~ms long, $786$-nm wavelength pump pulse hits the crystal $1.4$~mm below the surface. The surface tilt angle $\phi$ is measured for several values of the relative vertical positions $y$ of the probe and pump beams. Each curve is the result of $128$ averages. The filled areas correspond to the photothermal and piezo-orbital contributions derived from the fit (see Eq.~\ref{eq:combilineaire}).
(b) Space dependence $\Phi_i(y_i)$ (filled circles) of the piezo-orbital contribution to the measured tilt angle as obtained from the fit by Eq.~\ref{eq:combilineaire} (error bars are given by the fitting uncertainty on the parameter). A solid line corresponding to a Lorentzian surface displacement reproduces well the data. (c) Piezo-orbital Lorentzian-shaped maximum surface deformation inferred from the measured angles (not to scale).}
\label{fig:FitSimuVsY}
\end{figure*}

The principle of using time-resolved tomography for separating the conservative and dissipative contributions to the measured surface deformation is depicted in Figure~\ref{fig:unecourbe}(b) and (d). The time evolution of two photodeflection signals, acquired well apart on the crystal surface, are shown. When probed $350~\mu$m away from the pumping region (located at $y=0$), the surface tilt angle exhibits a damped response, with a delay significantly larger than the pump pulse duration. This slow response strongly evokes the heat diffusion process originating from a photothermal effect.
When the surface is probed closer to the pumping position ($y=-90\,\mu$m), a steep, stronger response is observed, followed by a decay comparable with the \levelm state lifetime.
This is compatible with the expected piezo-orbital response, synchronized with the population dynamics.

We repeat this measurement while scanning the position $y$ of the probe beam around the pumping position. This  enables point-by-point reconstruction of the surface shape, with a spatial resolution given by the probe beam waist on the crystal surface.
The experimental data [Figure~\ref{fig:FitSimuVsY}(a)] show the buildup and decay of a localized surface displacement, initially contained in a narrow region ($|y|<200~\mu$m), and progressively spreading after a few ms.

The measured tilts $\phi_{exp}(y_i,t)$  at positions $(y_i)$  are fitted to the following expression:
\begin{equation}
     \phi_{exp}(y_i,t)=\Phi_i(y_i)\ g_m(t) + \eta  \ \phi_{\textrm{therm}}(y_i,t)
   \label{eq:combilineaire}
\end{equation}
The first term corresponds to the piezo-orbital backaction, whose time dependence $g_m(t)$ is imposed by the atomic population evolution and whose space dependence $\Phi_i(y_i)$ is left as a set of adjustable parameters. The second term is the photothermal contribution, whose temporal and spatial dynamics are not separable due to the heat diffusion mechanism. For this contribution the factor $\eta$ is the only adjustable parameter. The two functions $g_m(t)$ and $\phi_\mathrm{therm}(y,t)$ are inferred from a theoretical model developed specifically for this geometry and described in detail in the SI.
The result of the fit is depicted as filled area plots in Figure~\ref{fig:FitSimuVsY}(a). The good agreement with the experimental data unequivocally demonstrates that the photothermal backaction alone cannot explain the observed behaviour: indeed, far from the pump beam position ($|y|>300~\mu$m), the photothermal contribution is the only contribution, which fully determines the fitting parameter $\eta$. Close to the center, the piezo-orbital contribution accounts for the difference between the experimental data and the simulated photothermal effect. We observe that in this region, both contributions are similar in size.
Figure~\ref{fig:FitSimuVsY}(b) shows the values of $(\Phi_i)$ describing the piezo-orbital contribution, as obtained from Eq.~\ref{eq:combilineaire}. The data are well described by a Lorentzian-shaped surface displacement with a $280~\mu$m FWHM and a $230$~pm amplitude at the end of the pumping pulse. The photothermal component is slightly underestimated in our model but captures the proper order of magnitude ($\eta=1.9$).

\begin{figure}[t]
  \centering
  \includegraphics[width=8.0cm]{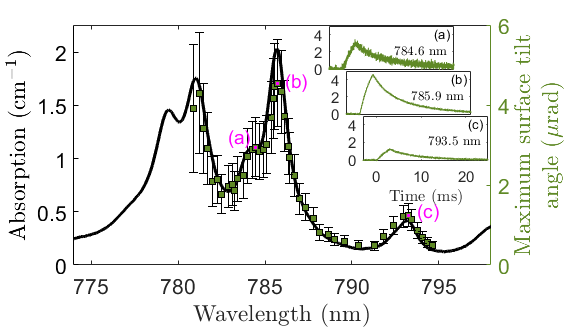}
  \caption{Variation of the maximum surface tilt angle with the pump laser wavelength (squares). The angle is normalized with respect to the pump power variations. On the same graph we plot the absorption coefficient of a $2$\%-doped Tm:YAG at room temperature (solid line)~\protect\cite{SciMatTmYAG}. The three insets (a, b, c) illustrate the similar time dependence of the  angle for three wavelengths.}\label{fig:vsLambda}
\end{figure}

In Figure~\ref{fig:vsLambda} we study the dependence of the photodeflection angle with respect to the pump laser wavelength, over an interval covering four partially overlapping absorption lines between the Stark multiplets \levelg and \levele~\cite{gruber1989spectra}. The value of $y=-90~\mu$m is chosen such that the photodeflection signal is the strongest. We observe that the maximum photodeflection angle closely follows the absorption profile, ruling out the possibility of a non-resonant contribution. It is also noticeable that the temporal variation of the photodeflection angle is unchanged over the whole wavelength range [Fig.~\ref{fig:vsLambda}(a,b,c)], indicating a preserved link between the optomechanical backaction and the internal atomic dynamics.

\section{Discussion}
\paragraph{About the piezo-orbital backaction}

Using the conservative nature of the piezo-orbital backaction, and given the  experiment geometry, the expected surface displacement $\Delta x$ due to the excitation of $N$ ions can be written:
\begin{equation}
\Delta x=N \hbar \frac{\Gstress}{w_p L},
\end{equation}
where $\Gstress =\frac{\partial \omega}{\partial \sigma}$ is the pressure sensitivity of the Tm atomic line, $w_p$ is the pump beam waist and $L$ is the crystal thickness along the pump laser propagation direction  (see SI).
The exact value of $\Gstress$ is not known for the \levelg$\rightarrow$\levelm transition in Tm:YAG. 
From the measured $\Delta x=230$~pm displacement we estimate it to $\Gstress\simeq 2\pi\times 80$~Hz/Pa, compatible with the typical values observed in a broad range of host crystals, rare-earth dopants, and transitions~\cite{louchet2019piezospectroscopic}.

Additional insight can be drawn by quantitatively estimating the force associated with the measured maximal piezo-orbital displacement, $\Fpo=EL \Delta x\simeq 0.27\,$N, where $E$ is the Young modulus of the crystal (see SI). This force, obtained with $230\,$mW pump power, is $8$ orders of magnitude larger than the radiation pressure force exerted by a similar laser beam that would be perfectly reflected on the surface of the crystal. We note that the force contribution from a single ion $\Fpo^0=\Fpo/N \simeq 700\,$aN is about $10$ times larger than the estimated piezo-excitonic backaction force from a single quantum dot~\cite{kettler2021inducing}, which may partly result from geometry constraints at the nanoscale~\cite{yeo2014strain}.

Interestingly, by comparing the experimental piezo-orbitally-driven surface displacement [Figure~\ref{fig:FitSimuVsY}(b)] with the one predicted by our model (see SI), we can estimate the fractional expansion $\xi_\textrm{exp}$ of a single Tm ion when promoted to the \levelm state, assuming a simple ionic radius dilation. We obtain $\xi_\textrm{exp} \simeq 2\cdot 10^{-4}$. Our optomechanical setup thus provides a unique access to this parameter, obtained by only considering the observed macroscopic displacement field without any exact knowledge of the microscopic orbital shape change, its propagation to the crystal field, or the coupling between the latter and the strain field in the crystal.

\paragraph{Perspectives}

An important aspect of this work relates to the comprehensive characterization of photothermal effects, which is generally difficult in hybrid systems, especially in micro- and nano-structures for which thermal actuation processes largely depend on defects and asymmetries~\cite{usami2012optical,kettler2021inducing}, and therefore remain largely unpredictable. Our time-resolved tomography scheme sheds new light on photothermal motion as a fundamental hybrid optomechanical process. This dissipative backaction contribution is directly linked to the non-radiative relaxation mechanisms of the quantum emitter, thereby providing a novel access to the internal atomic dynamics.

Moving further towards quantum hybrid optomechanics may be achieved by addressing a narrow spectral subset of ions via spectral hole burning (SHB) at cryogenic temperatures, where coherent quantum state manipulation becomes possible~\cite{rippe2008experimental}.
In this regime, the relative effect of strain on such narrow optical lines is increased by several orders of magnitude. These result in fluctuations of their atomic transition frequencies whose readout provides an access to the vibrations of the crystal~\cite{louchet2019piezospectroscopic}. The subsequent effect of the piezo-orbital backaction is to shift the atomic line, to first order. In our experiment, this frequency shift amounts to $\Delta\nu_\mathrm{po}= \frac{\Gstress }{2\pi}\frac{E}{w_{\mathrm{p}}}\Delta x \simeq 250\,\mathrm{MHz}$ (see SI), which is small relative to the absorption linewidth at room temperature ($\sim700$\,GHz). Concurrently, in typical SHB conditions, this shift is expected to be comparable or greater than the homogeneous linewidth. This suggests that the piezo-orbital backaction may represent a fundamental contribution in excitation-induced line broadening mechanisms, in particular towards elucidating the phenomenon of instantaneous spectral diffusion~\cite{louchet2023strain}.

Finally, rare-earth ion-doped crystals offer an unprecedented level of control that may be very attractive in the field of optomechanics. Indeed, in these materials, the relaxation mechanisms strongly depend on the transitions addressed, the choice of the ion amongst many possible atomic species~\cite{thiel2011rare} and its direct environment (host matrix, codoping)~\cite{zhang2020tailoring}. Besides, using volumic nanostructuration has been shown to alter the phonon spectral distribution and corresponding atomic dynamics~\cite{lutz2016modification}. All these degrees of freedom can be exploited to modify the thermal behaviour of the material, and, in turn, tune its optomechanical response, to some extent.
In particular, Yb-doped YLiF$_4$ represents a key material thanks to its minimal number of non-radiative decay channels. This property, already exploited in the context of laser cooling of solids~\cite{melgaard2016solid}, should also make this material an interesting candidate for conservative optomechanics.

\section{Conclusion}
In this work we have considered a hybrid optomechanical system composed of an ensemble of identical rare earth ions embedded in a bulk crystal, and coupled to internal strain. We have evidenced the generation of a mechanical response within this  system under resonant atomic excitation, which we find to be driven by two fundamental contributions: First, a conservative backaction force coined as \emph{piezo-orbital}, appearing with the promotion of the atoms to an excited state. This effect is expected in all impurity-doped crystals and reflects the interplay between the ions' orbital shape and the crystal field in their host matrix. The second contribution corresponds to a dissipative, photothermal backaction force associated with the non-radiative relaxation dynamics of the rare-earth ions. The resonant nature of the two effects and their fundamental connection to the internal atomic state dynamics were quantitatively confirmed.

This work reports for the first time a general method enabling to separately address these two fundamental mechanisms. Our approach sheds new light on hybrid systems and on the comprehension of non-conservative relaxation mechanisms to determine their associated fundamental quantum limits.
Our results were obtained at room temperature, but could in principle be transposed to cryogenic temperatures, where rare-earth ion-doped crystals exhibit exceptional coherence properties, setting interesting perspectives for quantum hybrid optomechanics and quantum technologies.

\begin{acknowledgments}
The authors are indebted to Jean-Louis Le Gouët and Ludovic Bellon for helpful discussions.
The authors acknowledge support from the French National Research Agency (ANR) projects ATRAP (ANR-19-CE24-0008), and MIRESPIN (ANR-19-CE47-0011).
This work has received support under the program "Investissements d’Avenir" launched by the French Government.

\end{acknowledgments}

%

\end{document}


\newcommand{\levele}{$^3H_4$\xspace}
\newcommand{\levelh}{$^3H_5$\xspace}
\newcommand{\levelm}{$^3F_4$\xspace}
\newcommand{\levelg}{$^3H_6$\xspace}
\newcommand{\Fpo}{F_{\mathrm{po}}}
\newcommand{\Fth}{F_{\mathrm{th}}}
\newcommand{\SFFth}{S_{\mathrm{FF}}^{\mathrm{th}}}
\newcommand{\SFFpo}{S_{\mathrm{FF}}^{\mathrm{po}}}
\newcommand{\SFFthres}{S_{\mathrm{FF}}^{\mathrm{th,res}}}
\newcommand{\perroothz}{\,\mathrm{Hz}^{-1/2}}
\newcommand{\nact}{n_{\mathrm{active}}}
\newcommand{\NTm}{N_{\mathrm{Tm}}}

\newcommand{\TC}[1]{{\color{green}{TC: #1}}}
\newcommand{\ALC}[2]{{\color{cyan}{\textbf{ALC: #1}} #2}}
\newcommand{\PV}[2]{{\color{magenta}{\textbf{PV: #1}} #2}}

\renewcommand{\thefigure}{S\arabic{figure}}
\renewcommand{\thesection}{\Roman{section}}

\title{Piezo-orbital backaction force in a rare-earth doped crystal:\\
Supplementary information}

\author{A. Louchet-Chauvet}
\email{anne.louchet-chauvet@espci.fr}
\affiliation{ESPCI Paris, Université PSL, CNRS, Institut Langevin, 75005 Paris, France}
\affiliation{Université Paris-Saclay, CNRS, Laboratoire Aim\'e Cotton, 91405 Orsay, France}

\author{P. Verlot}
\affiliation{Université Paris-Saclay, CNRS, ENS Paris-Saclay, CentraleSupélec, LuMIn, 91405, Orsay,
France}
\affiliation{Institut Universitaire de France, 1 rue Descartes, 75231 Paris, France}

\author{J.-P. Poizat}
\affiliation{Univ. Grenoble Alpes, CNRS, Grenoble INP, Institut Néel, 38000 Grenoble, France}

\author{T. Chanelière}
\affiliation{Univ. Grenoble Alpes, CNRS, Grenoble INP, Institut Néel, 38000 Grenoble, France}

\maketitle

\section{Tm:YAG level structure at room temperature}
\label{sec:tmyag}

One remarkable advantage of rare-earth ion-doped crystals (REICs) is the possibility to address a large ensemble of identical ions, which altogether behave as a single quantum degree of freedom~\cite{sun2000symmetry}. This enables to strongly enhance the optomechanical interaction, while fully preserving its quantum hybrid nature. At room temperature, this enhancement is considerable, all the ions of the crystal being virtually equivalent, due to the thermal line broadening that conceals the inhomogeneous broadening.

The spectroscopic properties of room-temperature REIC are well documented, motivated by the development of solid-state lasers~\cite{moncorge2005current}. Tm:YAG is no exception~\cite{caird1975characteristics} thanks to its central role in the development of $2~\mu$m lasers~\cite{stoneman1990efficient}. For the same reason, the quality of Czochralski-grown oxides has made significant progress over the past 40 years and the quality of Tm:YAG monocrystals is now very good~\cite{brandle2004czochralski} precisely driven by the need of laser materials with low parasitic losses to avoid damages.

\begin{figure*}[ht]
  \centering \includegraphics[width=16.5cm]{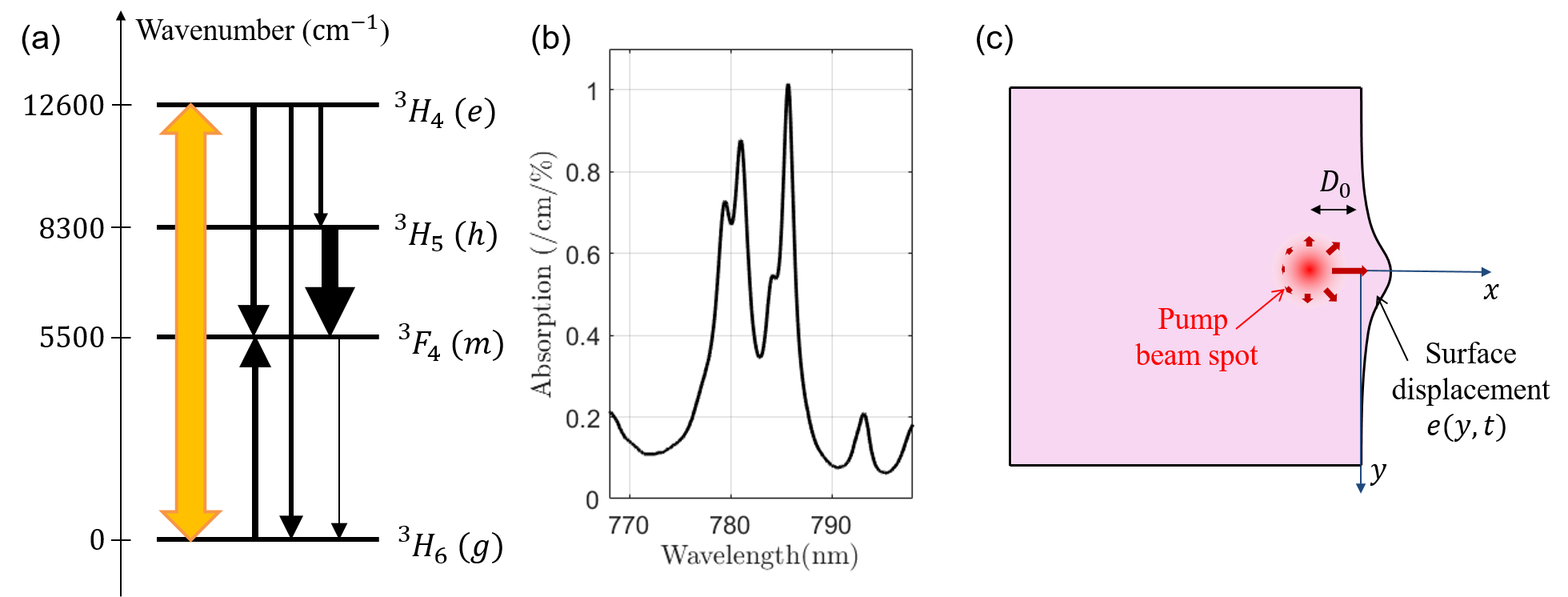}
  \caption{(a) Simplified Tm:YAG level system and main relaxation mechanisms. Each horizontal line represents a Stark multiplet, referred to as either its spectroscopic term such as \levele or a letter such as $e$. Optical excitation (double arrow) promotes Tm ions to the excited level. The  main relaxation mechanisms are depicted with single arrows. The thickness of the arrows symbolizes the magnitude of the relaxation rate. (b) Room-temperature absorption spectrum of Tm:YAG~\cite{SciMatTmYAG}. Each peak corresponds to a transition between a pair of \levelg and \levele Stark sublevels. (c) Schematic optomechanical transduction principle in a bulk crystal. The pumping beam is sent in the $z$ direction and focused close to an edge of the crystal. The optically induced volumic strain leads to a displacement field oriented preferentially towards the surface (solid arrows).}
  \label{fig:niveaux1surface}
\end{figure*}

Let us focus on the Tm ion's four Stark multiplets that take part in the laser operation in Tm:YAG, namely \levelg, \levelm, \levelh and \levele, respectively denoted as $g$, $m$, $h$, and $e$ in the following. The four-level system is depicted in Figure~\ref{fig:niveaux1surface}(a), together with the room-temperature absorption spectrum around $780$~nm [Figure~\ref{fig:niveaux1surface}(b)]. The crystal exhibits well-resolved lines reflecting the Stark multiplicity of the electronic levels~\cite{gruber1989spectra}. The strongest lines are found at $781$ and $786$~nm, corresponding to transitions from the lowest two, almost equally populated, Stark sublevels of \levelg (see Table~\ref{tab:boltzmann}).
Pumping Tm ions along one of these lines promotes them to the excited state $e$, from which they essentially decay into state $m$, partly cascading through the short-lived state $h$. Note that when the doping concentration exceeds $1\%$, Tm-Tm cross-relaxation significantly contributes to populate level $m$~\cite{armagan1992excited}. This optical pumping scheme can lead to a sizable fraction of a Tm ion ensemble stored for several ms in the metastable state.

\begin{table}[ht]
\begin{tabular}{c c c}
\hline
\hline
Level & Energy (cm$^{-1}$) & Occupancy \\
\hline
\levelg$(0)$ & $0$ & 0.35\\
\levelg$(1)$ & $27$ & 0.31\\
\levelg$(2)$ & $215$ & 0.12\\
\levelg$(3)$ & $239$ & 0.11\\
\levelg$(4)$ & $247$ & 0.11\\
\hline
\end{tabular}
\caption{Position and occupancy of Stark multiplet sublevels in the ground state.}
\label{tab:boltzmann}
\end{table}

\section{Experimental details}
\label{sec:expdetail}
The sample used in this work is a commercial YAG (Y$_3$Al$_5$O$_{12}$) single crystal (Scientific Materials) where Tm$^{3+}$ ions substitute to $c_{\mathrm{Tm}}=2$\% of the Y$^{3+}$ ions. It is a $4\times10\times10$~mm cuboid, polished on all 6 faces. The crystal sits on a piece of aluminum, and contacted with a thin layer of thermal paste. The rest of the crystal is in contact with air.
A $230$ mW pump beam is focused on a $w_p=22~\mu$m waist radius spot inside the crystal, entering the $10\times10$ surface of the crystal with normal incidence. The associated depth of focus is $4$~mm, matching the crystal length $L$ along the pump beam propagation axis. 

The number of thulium ions in the pump beam volume is about $n_Y c_\mathrm{Tm} \pi w_p^2 L=1.7\cdot10^{15}$ where $n_Y=1.38\cdot 10^{28}$m$^{-3}$ is the volumic density of yttrium ions in a pure YAG matrix. Each electronic level is split into several sublevels due to the crystal field in the YAG matrix. At room temperature, the thulium ions are distributed among the ground state \levelg multiplet following a Boltzmann distribution (see Table~\ref{tab:boltzmann}), leading to an effective number of ions addressed by the pump beam around $N_\mathrm{Tm}=5.9\cdot 10^{14}$ and $5.2\cdot 10^{14}$ ions in the lowest and second lowest Stark sublevel, respectively.

The pump beam is temporally chopped into $T_P=3$~ms-long rectangular pulses sent with a $86$~ms period, using an acousto-optic modulator. The transmitted pump beam intensity is initially attenuated by a factor $e^{-\alpha L}\simeq 0.5$ according to the Beer-Lambert-Bouguer law, where $\alpha = 2$~cm$^{-1}$ is the absorption coefficient at $786$~nm.
The transmitted intensity (see Figure~2(c) in main text) slowly increases during the pulse, revealing the effective optical pumping from the ground to the metastable state $m$.

The pump beam is generated by a Toptica DLPro extended cavity diode laser, injected into a  tapered amplifier. The laser is coarsely tunable over tens of nm around $785$~nm. The amplifier, on the other hand, is optimized for $793$~nm operation and its $3$~dB bandwidth is limited to a few nm. Therefore the optical power at the output of the amplifier drops by a factor of $\sim5$ when the wavelength is varied from $793$ to $781$~nm.

The probe beam is provided by a Nettest Tunics external cavity diode laser emitting a few mW at a completely different color ($1550$~nm), focused on the crystal distorted surface with a $60~\mu$m waist radius. The reflected beam is then collimated and directed towards the center of a Thorlabs PDQ30C quadrant photodiode with $150$~kHz bandwidth.
The quadrant photodiode signal is filtered via a DC-block and averaged over 128 shots to filter out slow technical noise due for example to air turbulence in the lab.

In the time-resolved tomography experiment (main text, Figure~3), the relative height of the pump and probe beams ($y$) is varied by vertically shifting the lens focusing the pump beam on the crystal, with the help of a manual linear translation stage.

\section{Modeling the optomechanical backaction}

In this section we first describe the atomic population evolution under resonant laser excitation, using a simple rate equation formalism. Then we present how we model the piezo-orbital and photothermal backaction mechanisms associated to this population evolution, in our specific bulk geometry.

\subsection{Rate equation modeling of the population evolution in Tm:YAG}
\label{sec:rateeq}
\begin{figure}[t]
  \centering
  \includegraphics[width=8.5cm]{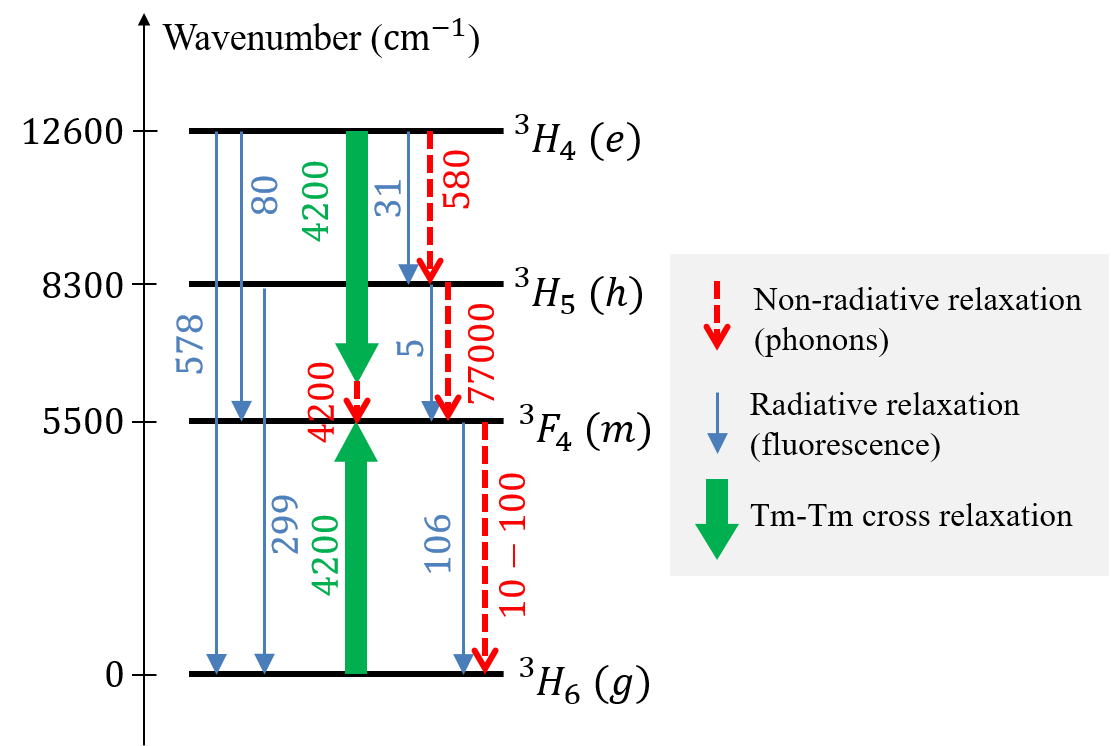}
  \caption{Relaxation mechanisms in the Tm ion level system. The vertically oriented numbers correspond to the relaxation rates (in s$^{-1}$) inferred from references \protect\cite{caird1975characteristics} and \protect\cite{guillemot2020watt}, assuming a $2\%$ Tm doping concentration. The non-radiative relaxation rate from $m$ to $g$ according to \protect\cite{caird1975characteristics} is $10$~s$^{-1}$, but we extend it to $100$~s$^{-1}$ in our model so that the metastable state lifetime matches the one measured in our sample ($5$~ms).}
  \label{fig:niveaux2}
\end{figure}

The complete relaxation mechanisms at room temperature in the Tm ion's four-level system are represented in Figure~\ref{fig:niveaux2}.
The excitation along the $g\rightarrow e$ transition is represented by a time- and space-dependent pumping rate $R$, proportional to first order to the optical irradiance of the pump laser. The pumping rate is derived experimentally from the measurement of the transmitted pump power (see Figure~2(c) in the main text). As an example, a $230$~mW pump pulse focused on a $22~\mu$m waist radius corresponds to a $R=340$~s$^{-1}$ pumping rate along the $g\rightarrow e$ transition during the pumping pulse in the center of the Gaussian beam. We model the population dynamics with a rate equation model, leading to the following differential equation system:

\begin{widetext}
\begin{equation*}
\frac{d}{dt}\left(\begin{matrix}
n_g \\
n_m\\
n_h\\
n_e
\end{matrix}\right)
= \left(\begin{matrix}
-R & \Gamma_{mg}  & \Gamma_{hg} & R+\Gamma_{eg} - W_{CR}\\
0 & -\Gamma_{mg} & \Gamma_{hm} & \Gamma_{em}+2W_{CR}\\
0 & 0 & -\Gamma_{hm}-\Gamma_{hg} & \Gamma_{eh}\\
R & 0 & 0 & -R-\Gamma_{em}-\Gamma_{eg}-\Gamma_{eh}-W_{CR}
\end{matrix}\right)
\left(\begin{matrix}
n_g \\
n_m\\
n_h\\
n_e
\end{matrix}\right)
\label{eq:rateeq}
\end{equation*}
\end{widetext}
and the populations verify $n_g+n_m+n_h+n_e=1$.
The notations are as follows: $\Gamma_{ij}$ is the total relaxation rate from level $i$ to $j$, including radiative and non-radiative channels.
$W_{CR}$ is the concentration-dependent cross-relaxation rate.
Solving this differential system yields the time- and space-dependent atomic populations in all four levels.

\subsection{Piezo-orbital surface displacement}
\label{sec:piezomodel}

\begin{figure}
  \centering
  \includegraphics[width=8.0cm]{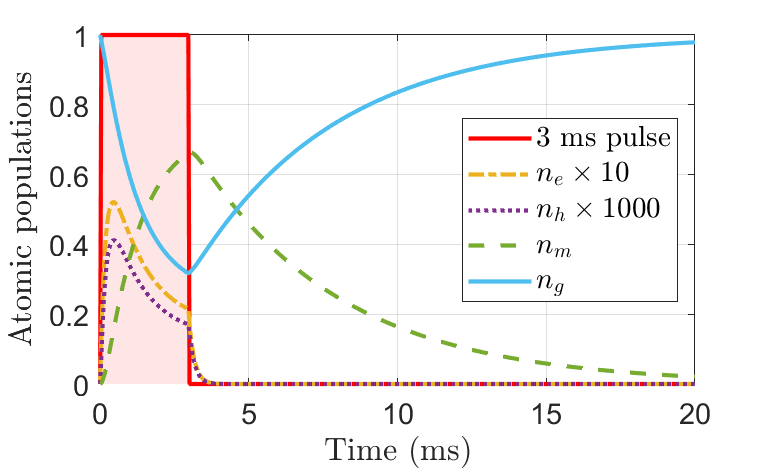}
  \includegraphics[width=7.5cm]{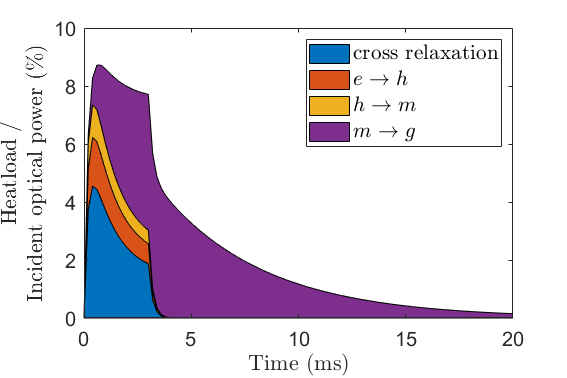}
  \caption{(a) Atomic populations in a room-temperature Tm:YAG 4-level system, at the center of a $230$~mW pumping pulse focused on a $22~\mu$m waist spot. The pulse is rectangular and $3$~ms long. (b) Solid line: calculated heatload due to phonon emission triggered by a $3$-ms rectangular pulse in a Tm:YAG crystal. The heatload power is normalized with respect to the incident power ($230$~mW in this calculation). The heatload is calculated at the center of the pump beam and broken down into its 4 contributions according to Eq.~\ref{eq:phonons}.}
  \label{fig:populations_phonons}
\end{figure}

The spatio-temporal variations of the atomic populations excited by a pulsed pump beam are given by the rate-equation formalism described above. Considering the Gaussian pump beam profile, we obtain the space and time-dependent atomic populations dynamics in the four Tm levels. In Figure~\ref{fig:populations_phonons}(a) we plot the atomic populations at the center of the beam and observe that more than $60$~\% of the population reaches the metastable state in the center of the beam at the end of the pumping pulse.

Since the populations in levels $e$ and $h$ never exceed $5\%$ in the center of the beam, we only focus on the piezo-orbital backaction triggered by the population transfer between levels $g$ and $m$.
For simplicity, we consider that the effect of this population transfer is a mere change of the ionic radius, defined as $\xi=(r_m-r_g)/r_g$ where $r_m,r_g$ refer to the ionic radii in the metastable and ground states respectively.
We focus on the fraction of the displacement field exclusively oriented in the $x$ direction towards the nearby crystal surface at $x=0$. We estimate the displacement $e_{\textrm{piezo}}$ of the surface as the sum of the local displacements along the $x$ direction:
\begin{equation}
e_{\textrm{piezo}}(y,t)= \sqrt[3]{n_{\textrm{Tm}}}\ 2r_{g} \xi \int_{-\infty}^0  n_m(x,y,t) dx
\label{eq:epiezo1}
\end{equation}
where $n_{\textrm{Tm}}$ is the volumic density of Tm ions in the crystal, and $n_m$ is the fractional population in the $m$ state at position $(x,y)$ and time $t$. The ionic radius of Tm$^{3+}$ in the ground state ($r_g=0.994$\AA) is found in~\cite{shannon1976revised}.
Equation~\ref{eq:epiezo1} reflects the instantaneous nature of the piezo-orbital backaction and its direct connection to the atomic population in the metastable state.

Given the values of the maximum pumping rate $R=340$~s$^{-1}$ at the center of the pumping beam and of the pump pulse duration $T_P=3$~ms, we have $R T_P\lesssim 1$. Therefore we consider the population to be linearly dependent on the laser power. Consequently the atomic populations can be written as the product of two separable functions of time and space. For example the metastable population reads as: $n_m(x,y,t)=n_m(x,y,T_P)g_m(t)$ where $g_m(t)=\frac{n_m(0,0,t)}{n_m(0,0,T_P)}$.

Using this property for the metastable population, Equation~\ref{eq:epiezo1} simplifies to
\begin{equation}
e_{\textrm{piezo}}(y,t)=e_{\textrm{piezo}}(y,T_P) g_m(t)
\label{eq:epiezo2}
\end{equation}

\subsection{Photothermal surface displacement}
\label{sec:thermomodel}
We now address the mechanical deformation associated with the heating contribution of the pumping pulse linked to the phonons emitted throughout the atomic relaxation process. The modeling is done in two steps. First, we use the population dynamics to estimate the non-radiative phonon contribution to heat using the parameters documented in the laser literature. Second, we solve the associated heat equation to obtain a thermal map of the crystal in space and time.

\subsubsection{Phonon-driven heatload}
According to Figure~\ref{fig:niveaux2}, each step of the decay of the Tm ions from the excited to the ground state is partly non-radiative and generates heat. In addition, in the cross-relaxation mechanism, phonons are produced to compensate for the detuning between the $g\rightarrow m$ and the $e \rightarrow m$ transitions~\cite{armagan1992excited}.
Overall, the total heatload power reads as the sum of four contributions:
\begin{equation}
P_{\textrm{heat}}(r,t)=N_{\textrm{Tm}} \left[P_{CR}(r,t) + P_{eh}(r,t) + P_{hm}(r,t) + P_{mg}(r,t) \right] \label{eq:phonons}
\end{equation}
where $N_{\textrm{Tm}}$ is the number of atoms addressed by the pump beam in the illuminated volume. Each term of the sum can be expressed in terms of atomic populations:
\begin{eqnarray}
P_{CR}(r,t) &=&W_{CR} \Delta_{CR} n_e(r,t) \nonumber \\
P_{eh}(r,t) &=& \Gamma_{eh}^{NR} \Delta_{eh} n_e(r,t)\nonumber\\
P_{hm}(r,t) &=& \Gamma_{hm}^{NR} \Delta_{hm} n_h(r,t) \\
P_{mg}(r,t) &=& \Gamma_{mg}^{NR} \Delta_{mg} n_m(r,t) \nonumber
\end{eqnarray}
where $\Delta_{CR}$ is the energy detuning between the two transitions involved in the cross-relaxation process, $\Delta_{ij}$ is the energy splitting between levels $i$ and $j$.
$\Gamma_{ij}^{NR}$ is the non-radiative part of the relaxation rate along transition $i\rightarrow j$. Each contribution to the heatload therefore exhibits a time dependence related to the population evolution in the 4-level system, and a space dependence originating from the pump laser Gaussian profile. Note that we do not include radiation trapping~\cite{molisch1998radiation} for photons emitted from the metastable state, because this mechanism occurs in the whole crystal volume, given the moderate absorption coefficient along the $g\rightarrow m$ transition~\cite{stoneman1990efficient}.

The total calculated heatload $P_{\mathrm{heat}}(r=0,t)$ in the center of the beam is shown in Figure~\ref{fig:populations_phonons}(b). Interestingly, the total heatload  rapidly grows during the pulse and slowly decays after the end of the pulse. This persistence originates from the phonon generation along the $m\rightarrow g$ decay path. This contribution is proportional to the metastable state population $n_m(t)$, which decays within a few ms.

The shortest timescale of the heatload variation is around $\tau_{\textrm{heat}}=1$~ms, leading to a corresponding thermal diffusion length $L_{\textrm{diff}}= \sqrt{\alpha_{\textrm{diff}}\tau_{\textrm{heat}}} \simeq  60~\mu$m (where $\alpha_{\textrm{diff}}=4\times 10^{-6}$~m$^2.$s$^{-1}$ is the thermal diffusivity in YAG ~\cite{klein1967thermal}). In a sample smaller than $L_{\textrm{diff}}$, the diffusion of such a heatload would be considered instantaneous. In a bulk monocrystal with millimetric dimensions however, the diffusion mechanism becomes prominent and must therefore be taken into account for predicting the time-dependent temperature distribution. 

\subsubsection{Heat equation}

Given the problem geometry, we write the heat equation in cylindrical coordinates, assuming translational invariance along the cylinder axis (pump laser beam) as a simplification, and rotational invariance around its axis:
\begin{widetext}
\begin{equation}
\frac{\partial T}{\partial t}=
\alpha_{\textrm{diff}} \frac 1 r \frac{\partial}{\partial r}\left(r\frac{\partial T}{\partial r}\right)+ \frac{q(r,t)} {\rho c_p}
\label{eq:EqChaleur}
\end{equation}
\end{widetext}
The temperature at the start of the pumping pulse is assumed uniform and equal to zero. The source term $q(r,t)$ is the time- and space-dependent volumic heatload derived from $P_{\textrm{heat}}(r,t)$ (see Eq.~\ref{eq:phonons}).

The presence of the nearby surface alters the heat diffusion mechanism. Indeed, the low air conductivity imposes a zero heat flow through the surface. To account for this boundary condition we make the problem symmetric by assuming a second identical cylindrical heat source inside the infinite crystal centered at position $(D_0,0)$.

Solving the heat equation yields the temperature distribution in the crystal volume $T(x,y,t)$.
The thermal expansion of the crystal in any direction is given by $\kappa T$ where $\kappa=8\times10^{-6}$~K$^{-1}$ is the YAG linear thermal expansion coefficient.
Finally, considering again only the displacement field directed towards the nearby surface at $x=0$, the displacement of the crystal surface due to the photothermal effect reads as:
\begin{equation}
e_{\textrm{therm}}(y,t)=\kappa  \int_{-\infty} ^0 dx \  T(x,y,t)
\label{eq:ethermo}
\end{equation}
To sum up, besides being deeply linked to the internal dynamics of the Tm ions via the source term of the heat equation, the photothermal effect also reflects the diffusion mechanism that occurs in a bulk sample through the time- and space-dependent temperature distribution.

\subsection{Surface displacement and photodeflection signal}
\label{sec:surfacemodel}

\begin{figure}[t]
\centering
\includegraphics[width=8.6cm]{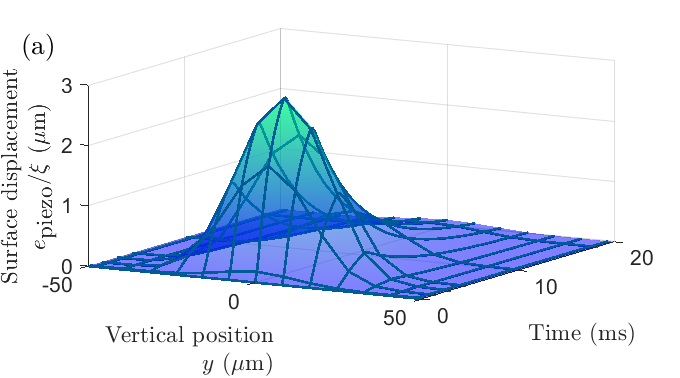}
\includegraphics[width=8.6cm]{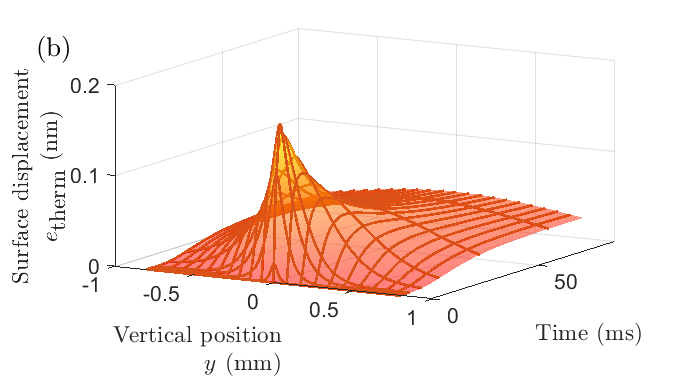}
\caption{Surface displacement for the (a) piezo-orbital and (b) photothermal backactions as predicted by our two models, assuming a $3$~ms pumping pulse. Note the different scales for the two graphs. For the piezo-orbital backaction we represent the surface displacement normalized by the fractional ionic radius change $\xi$.}
\label{fig:eSimul}
\end{figure}

\begin{figure}[t]
\centering
\includegraphics[width=8.6cm]{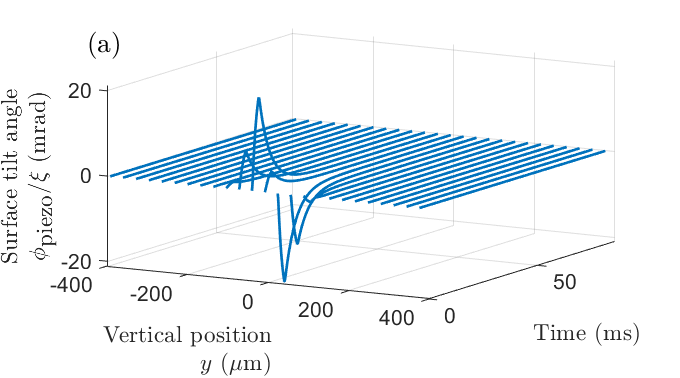}
\includegraphics[width=8.6cm]{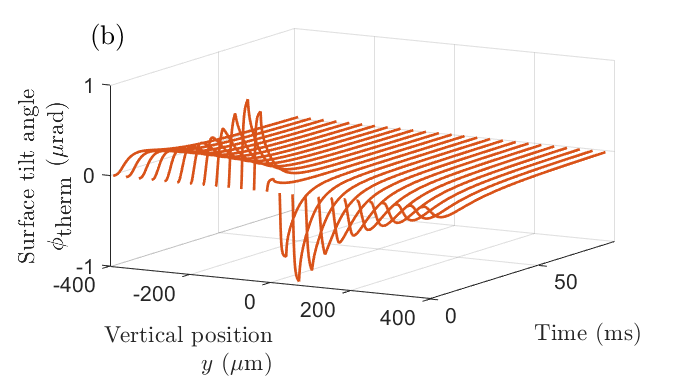}
\caption{Deflection angles $\phi(y,t)$ derived from our piezo-orbital (a) and photothermal (b) models, including the effect of the probe beam Gaussian profile.}
\label{fig:thetaSimul}
\end{figure}

In Figure~\ref{fig:eSimul} we plot the expected surface displacements $e_{\textrm{piezo}}(y,t)$ and $e_{\textrm{therm}}(y,t)$ provided by our two models described in sections~\ref{sec:piezomodel} and \ref{sec:thermomodel}. Both effects give rise to a bump on the crystal surface, centered in $y=0$. In the piezo-orbital case, the bump exhibits a constant spatial width, while its time dependence is given by the metastable population evolution $g_m(t)$ (see Eq.~\ref{eq:epiezo2}).
The photothermal contribution, on the other hand, exhibits a  progressive spatial broadening due to heat diffusion. In addition, its time evolution strongly depends on the value of $y$: the photothermal response is quasi-instantaneous on the center of the bump ($y=0$), closely resembling the piezo-orbital response, but it is manifestly delayed away from the bump.
The corresponding surface tilt angles $\phi_\textrm{piezo}(y,t)$ and $\phi_\textrm{therm}(y,t)$ are then obtained by from the derivative of the displacements $e_{\textrm{piezo}}(y,t)$ and $e_{\textrm{therm}}(y,t)$ along the $y$ dimension. To account for the finite size of the probe beam on the surface, we convolve the simulated angles with the Gaussian profile of the probe beam. The results are shown in Figure~\ref{fig:thetaSimul}. Again, the photothermal effect reveals the diffusion process inherent to heating mechanisms, whereas the piezo-orbital contribution is more localized spatially and directly reflects the evolution of the metastable state population.

\section{Macroscopic estimation of the piezo-orbital effect magnitude}
\label{sec:macroscopic}
\begin{figure}[ht]
  \centering
  \includegraphics[width=5cm]{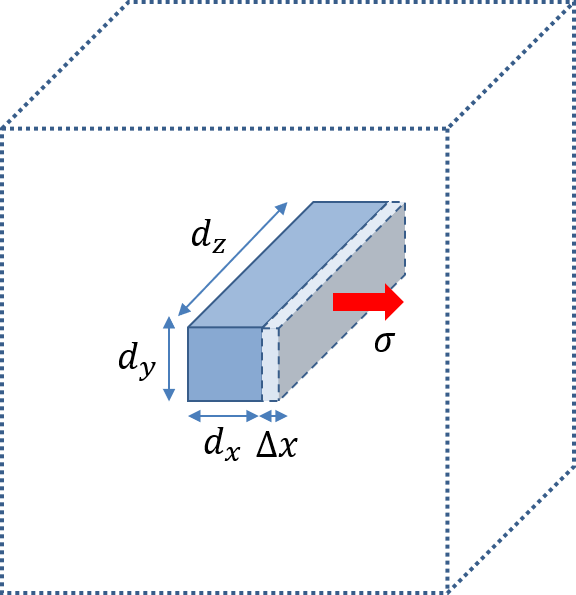}
  \caption{Notations used for the macroscopic estimation of the piezo-orbital magnitude in section~\ref{sec:macroscopic}.}\label{fig:cuboid}
\end{figure}

In this section we demonstrate that the magnitude of the piezo-orbital effect is fundamentally related to the pressure sensitivity of the atomic line $G_{\textrm{stress}}=\frac{\partial \omega}{\partial \sigma}$, expressed in Hz/Pa. We consider the mechanical system made of a $d_x\times d_y \times d_z$ rectangular cuboid (see Figure~\ref{fig:cuboid}). The piezo-orbital effect of $N$ excited ions within this volume creates a pressure $\sigma$ on the $yz$ surface that we can relate to the expansion $\Delta x$ along the $x$ dimension via Hooke's law:
\begin{equation}
\sigma =E\ \frac{\Delta x}{d_x}
\label{eq:pressure}
\end{equation}
where $E$ is the Young modulus of the medium.
The conservative piezo-orbital force resulting from this pressure in the $x$ direction reads as $\Fpo=\sigma \, d_y d_z$. Using Eq.~\ref{eq:pressure} we derive $k_m$, the mechanical stiffness of the material associated with the normal deformation of the crystal $yz$ surface, assuming $d_x=d_y$:
\begin{equation}
k_m=\frac{\Fpo}{\Delta x}=E\ d_z
\label{eq:stiffness}
\end{equation}
$\Fpo$ can also be written as the spatial derivative of the $N$ excited ions' internal energy and undergoing the volumic change~\cite{leech1965fundamental}.
\begin{equation}
\Fpo=\frac{\partial}{\partial x} \left(N \hbar \omega\right) = N \hbar \frac{\partial \omega}{\partial x}
\end{equation}
The factor $\frac{\partial \omega}{\partial x}$ is the atomic line sensitivity to a displacement of the $yz$ surface. Again, using Hooke's law, we obtain
\begin{equation}
\Fpo=N \hbar \frac E {d_x} G_{\textrm{stress}}
\label{eq:force}
\end{equation}
The number of ions is given by $N=\NTm\nact$  where $\NTm$ was estimated in Section~\ref{sec:expdetail} and $\nact\simeq0.6$ is  the fraction of ions in the volume that generate the volumic change (see Figure~\ref{fig:populations_phonons}). Relating pressure and force (equations~\ref{eq:pressure} and \ref{eq:force}) we obtain:
\begin{equation}
\Delta x=N \hbar \frac{G_{\textrm{stress}}}{d_y d_z}
\end{equation}

The pressure $\sigma$ in the cuboid volume due to the piezo-orbital backaction leads to a shift of the atomic resonance frequency, that we can relate to the piezo-orbital displacement using Hooke's law (Eq.~\ref{eq:pressure}):
\begin{equation}
\Delta\nu_\mathrm{po}= \frac{G_{\mathrm{stress}}}{2\pi} \sigma =  \frac{G_{\mathrm{stress}}}{2\pi} \frac{E}{d_x}\Delta x
\end{equation}

We apply these results to the geometry of our photodeflection experiment, using  $d_y=w_p$ and $d_z=L$, to obtain the relevant expressions used in the main text:
\begin{eqnarray}
\Delta x &=& N \hbar \frac{G_{\textrm{stress}}}{w_p L}\\
\Fpo &=& E L \Delta x\\
\Delta\nu_\mathrm{po} &=& \frac{G_{\mathrm{stress}}}{2\pi} \frac{E}{w_p}\Delta x
\end{eqnarray}

\bibliographystyle{unsrt}